\documentclass[rnote]{aa}
\usepackage{graphicx}
\usepackage{txfonts}
\begin{document}

   \title{Testing Einstein's Equivalence Principle with Cosmological Fast Radio Bursts behind Clusters of Galaxies}

   %\subtitle{anti-correlation between the X-ray variability amplitude and the luminosity}

   \author{
          Shuang-Nan Zhang\inst{1,}\inst{2}
          }

   %\offprints{G. Wuchterl}

   \institute{Key Laboratory of Particle Astrophysics, Institute of High Energy Physics,
                Chinese Academy of Sciences, Beijing 100049, China\\
              \email{zhangsn@ihep.ac.cn}
         \and
 National Astronomical Observatories, Chinese Academy Of Sciences, Beijing 100012, China\\               
 }
       
   \date{}

% \abstract{}{}{}{}{}
% 5 {} token are mandatory

  \abstract
  % context heading (optional)
  % {} leave it empty if necessary
   {}
  % aims heading (mandatory)
   {Recently, cosmological fast radio bursts (FRBs) have been used  to provide the most stringent limit up to date on Einstein's Equivalence Principle (EEP). We study how to further test EEP with FRBs.}
  % methods heading (mandatory)
   {Future systematic radio surveys will certainly find abundant FRBs at cosmological distances and some of them will inevitably be located behind clusters of galaxies.  Here we suggest to use those FRBs to further  test EEP.}
  % results heading (mandatory)
   {We find that the robustness and accuracy of testing EEP can be improved further by orders of magnitude with these FRBs. The same methodology can also be applied to any other types of fast and bright transients at cosmological distances.}
  % conclusions heading (optional), leave it empty if necessary
   {}

   \keywords{Gravitation }

   \authorrunning{Shuang-Nan Zhang}
   \titlerunning{EEP test with FRBs behind Clusters of Galaxies}
   \maketitle
%
%________________________________________________________________

Recently, cosmological fast radio bursts (FRBs)  (Lorimer et al. 2007; Keane et al. 2012; Thornton et al. 2013; Burke-Spolaor and Bannister 2014; Spitler et al. 2014; Masui et al. 2015; Petroff et al. 2015; Ravi et al. 2015) have been used by Wei et al. (2015) to provide the most stringent limit up to date on Einstein's Equivalence Principle (EEP), through the
relative differential variations at radio energies of one of the parametrized post-Newtonian (PPN) parameters,
i.e., the parameter $\gamma$ (see, e.g., Will 2006; Gao et al. 2015).  Specifically, assuming as a lower limit that the time delays of photons of a FRB with different energies
are caused mainly by the gravitational potential of the Milky Way (MW), Wei et al. (2015) obtained a strict upper limit
$\left[\gamma(1.23\; \rm GHz)-\gamma(1.45\; \rm GHz)\right]<4.36\times10^{-9}$.

Future systematic radio surveys will certainly find abundant FRBs at cosmological distances and some of them will inevitably be located behind clusters of galaxies (CGs).  Here we suggest that the accuracy of testing EEP can be improved further by orders of magnitude with these FRBs. Following Wei et al. (2015), the key principle of testing EEP with a FRB behind a CG is expressed by
\begin{equation}
\Delta t_{\rm CG}\equiv\Delta t_{\rm obs}-\Delta t_{\rm DM}>\frac{\gamma_{\rm 1}-\gamma_{\rm 2}}{c^3}\int~U(l)d l \ ,
\label{eq:deltatnew}
\end{equation}
where $\Delta t_{\rm obs}$ is the observed time delay between photons in two different energy bands of the FRB,
$\Delta t_{\rm DM}$ is the time delay contribution from the dispersion by the
line-of-sight (LOS) free electron content, $l$ is the LOS (comoving) distance of the photons of the FRB through the CG,  $U(l)$ is the gravitational potential of the CG.

For simplicity, the dark matter halo of a CG is taken as the NFW profile (Navarro et al. 1996), we therefore have
\begin{equation}
\int~U(l)d l=4\pi G M_{\rm CG}f(b,r_{\rm max}) \ ,
\end{equation}
where $G$ is the gravitational constant, $f(b, r_{\rm max})$ is a factor depending upon the specific values of the impact parameter $b$ of the FRB and the assumed maximum radius of the CG $r_{\rm max}$,  and $M_{\rm CG}$ is the total mass of the CG. If we take $r_{\rm max}$ as the virial radius of the CG and for $b$ ranging from zero to a fraction of $r_{\rm max}$, $f(b,r_{\rm max})\sim 1-10$. We therefore have,
\begin{equation}
\Delta \gamma_{\rm CG}^{<}\equiv(\gamma_{1}-\gamma_{2})_{\rm CG}<\frac{c^{3}\Delta t_{\rm CG}}{GM_{\rm CG}}\ .
\end{equation}
In comparison for a FEB passing through only MW \cite{Wei2015}, we have
\begin{equation}
\Delta \gamma_{\rm CG}^{<}\sim \frac{M_{\rm MW}}{M_{\rm CG}}\Delta \gamma_{\rm MG}^{<}  \ .
\end{equation}
Since $M_{\rm CG}\sim 10^{14}-10^{15}$ solar masses, we can further constrain EEP by two to three more orders of magnitudes with FRBs behind CGs, in comparison with using FRBs passing through only MW (Wei et al. 2015). The same test method can also be applied to PRBs behind any massive structures in the universe, allowing statistically robust tests of EEP.

We can also calculate the weighted averages  $\overline{\Delta t}_{\rm CG}$ and $\overline{\Delta t}_{\rm F}$ , for FRBs behind CGs or any other massive structures and field FRBs, respectively, and then further test EEP by,
\begin{equation}
\overline{\Delta \gamma}_{\rm CG}^{<}<\left(\overline{\Delta t}_{\rm CG}-\overline{\Delta t}_{\rm F}\right)\frac{c^{3}}{GM_{\rm CG}}\ .
\end{equation}
The weight $w$ for each FRB may be taken as $w=M_{\rm CG}/\sigma^2(t)$, where $\sigma(t)$ is the measurement error of $\left(\Delta t_{\rm obs}-\Delta t_{\rm DM}\right)$ for the FRB.

Clearly even more robust and stringent test of EEP can be achieved this way. Statistically significant detection of $\left(\overline{\Delta t}_{\rm CG}-\overline{\Delta t}_{\rm F}\right)\neq0$ would provide evidence for violation of EEP, independent of any model and specific properties of FRBs. The same methodology can also be applied to any other types of fast and bright transients at cosmological distances.

\begin{acknowledgements}
This work is partially supported by the National Basic Research Program (``973'' Program)
of China (Grants 2014CB845802), the National Natural Science
Foundation of China (grants No. 11373036  and 11133002), the Qianren start-up grant 292012312D1117210, and the Strategic Priority Research Program
``The Emergence of Cosmological Structures'' (Grant No. XDB09000000) of the Chinese Academy
of Sciences.
\end{acknowledgements}

\end{document}